\documentclass[aps,a4paper,twocolumn,floatfix]{revtex4}
\usepackage{times}
\usepackage{graphics}
\usepackage{epsfig,calc}

\usepackage{bm}

\usepackage{amsmath}

\def\AmS{{\protect\the\textfont2
        A\kern-.1667em\lower.5ex\hbox{M}\kern-.125emS}}

\citeindextrue

\newcommand{\eb}{\varepsilon_{\text{b}}}
\newcommand{\sbond}{s_{\text{b}}}

\newcommand{\bbond}{\mathbf{b}}

\newcommand{\rj}{\mathbf{R}_{j}}
\newcommand{\ri}{\mathbf{R}_{i}}

\newcommand{\f}{\mathbf{F}}
\newcommand{\df}{\delta \mathbf{F}}
\newcommand{\ttau}{\bm{\tau}}
\newcommand{\dtau}{\delta \bm{\tau}}
\newcommand{\oomega}{\bm{\omega}}
\newcommand{\gamr}{\gamma_{\text{r}}}
\newcommand{\Arot}{\mathbf{A}}

\newcommand{\rc}{r_{c}} 
\newcommand{\bthree}{\text{B}_3}
\newcommand{\bfour}{\text{B}_4}
\newcommand{\bfive}{\text{B}_5}

\newcommand{\tm}{\theta_{\text{m}}}
\newcommand{\pmax}{\phi_{\text{m}}}

\newcommand{\nb}{n_{\text{b}}}

\newcommand{\ccc}{\rho_{\text{cc}}}
\newcommand{\tobs}{t_{\text{obs}}}
\newcommand{\tf}{6\times 10^5}
\newcommand{\fc}{f_{\text{c}}}
\newcommand{\kt}{k_{\text{B}}T}
\newcommand{\kb}{k_{\text{B}}}
\newcommand{\conc}{C_0}
\newcommand{\ct}{\fc}
\newcommand{\nbe}{n_{\text{be}}}
\newcommand{\nfb}{b_{\text{net}}}
\newcommand{\bmult}{b_{\text{mult}}}
\newcommand{\bpls}{b_{+}}
\newcommand{\bmns}{b_{-}}

\begin{document}


\title{Dynamic Pathways for Viral Capsid Assembly}
\author{Michael F. Hagan}
\affiliation{Department of Chemistry, University of California, Berkeley, CA 94720}
\author{David Chandler\footnote{Corresponding author.}}
\affiliation{Department of Chemistry, University of California, Berkeley, CA 94720}

\date{\today}
\begin{abstract}We develop a class of models with which we simulate the assembly of particles into T1 capsid-like objects using Newtonian dynamics.  By simulating assembly for many different values of system parameters, we vary the forces that drive assembly.  For some ranges of parameters, assembly is facile, while for others, assembly is dynamically frustrated by kinetic traps corresponding to malformed or incompletely formed capsids.  Our simulations sample many independent trajectories at various capsomer concentrations, allowing for statistically meaningful conclusions. Depending on subunit (i.e., capsomer) geometries, successful assembly proceeds by several mechanisms involving binding of intermediates of various sizes. We discuss the relationship between these mechanisms and experimental evaluations of capsid assembly processes.
\end{abstract}
\maketitle

\section{Introduction}
\label{sec:intro}
This paper is devoted to introducing a simple class of capsomer models, and demonstrating that Newtonian dynamics of these models exhibit spontaneous assembly into 60-unit icosahedral capsids, depending upon conditions (i.e., particle concentration and force field parameters). We believe it is the first report of statistically meaningful simulations of capsid assembly that follow from unbiased dynamics obeying time-reversal symmetry and detailed balance.

The formation of viral capsids is a marvel of natural engineering and design. A large number (from 60 to thousands) of protein subunits assemble into complete, reproducible structures under a variety of conditions while avoiding kinetic and thermodynamic traps. Understanding the features of capsid components that enable such robust assembly could be important for the development of synthetic supra-nano assemblies. In addition, this knowledge is essential for the development of anti-viral drugs that inhibit capsid assembly or disassembly and could focus efforts to direct the making of highly specific drug delivery vehicles. These goals necessitate the ability to manipulate when and where capsids assemble and disassemble. Thus, we seek to determine what externally or internally controlled factors promote or alleviate dynamic frustration in the capsid assembly process. Although many viruses assemble with the aid of nucleic acids and scaffolding proteins, the first step toward this objective is to understand the inherent ability of subunit-subunit interactions to direct spontaneous assembly.

The equilibrium properties of viral capsids have been the subject of insightful theoretical investigations (e.g Refs. \onlinecite{Crick1956,Caspar1962,Bruinsma2003,Zandi2004,Twarock2004,Chen2005}) and the assembly process has been investigated in a number of experiments (e.g. Refs. \onlinecite{Fraenkel1955,Butler1978,Klug1999,Fox1994,Zlotnick1996,Zlotnick2000,Singh2003,Willits2003,Wu2004}), yet this process is still poorly understood for many viruses  (e.g. Ref. \onlinecite{Valegard1997}). Assembly is difficult to analyze experimentally because most intermediates are transient. With single molecule techniques, it is now possible to directly probe intermediate structures. Each intermediate, however, is a member of a large ensemble of structures and pathways that comprise the overall assembly process. Formation of an intermediate requires collective binding events that are regulated by a tightly balanced competition of forces between individual subunits. It is difficult, with experiments alone, to parse these interactions for the factors that are critical to the assembly process. Thus, it is useful to have complementary computational models in which the effects of different interactions can be isolated and monitored.

Studying assembly through computation is challenging because short range subunit-subunit properties regulate the formation of overall structure. Binding and unbinding rates of individual subunits are orders of magnitude faster than the overall assembly times. Furthermore, these rates are controlled by interactions defined on atomic lengths, which are three orders of magnitude smaller than typical capsid sizes.  Prior computational studies provide valuable insights that we build upon; in particular, Zlotnick pioneered a rate-equation description of assembly \cite{Zlotnick1994} and Berger and co-workers developed particle based methods \cite{Berger1994}.  Earlier studies, although an important foundation for our work, are limited in that they have been based upon pre-conceived pathways of assembly \cite{Zlotnick1994, Zlotnick1999,Endres2002,Reddy1998, Keef2005}, or dynamics that did not obey detailed balance \cite{Rapaport2004,Berger1994, Berger2000, Schwartz2000}, or dynamics that was anecdotal \cite{Schwartz1998, Rapaport1999, Rapaport2004, Zhang2004}.  These approaches can be useful and physical justifications for them can be made.  Nevertheless, we seek to avoid these limitations in order to understand the nature of possible kinetic traps and the extent of ensembles of successful assembly events.
	In Section \ref{sec:model}, we present our class of models for capsid subunits. We evaluate the thermodynamic properties of this model in Section \ref{sec:thermo}, and then discuss the results of dynamical simulations in Section \ref{sec:results}.  By simulating assembly for many different values of system parameters, we vary the strength of the forces that either drive or thwart assembly. We identify regions of parameter space in which two forms of kinetic traps prevail and we elucidate processes by which dynamic frustration is avoided in other regions of parameter space.

\section{Model}
\label{sec:model}
{\bf Capsomers.}  Capsid proteins typically have several hundred residues that fold into well defined shapes with specific interactions that lead to attractions between complementary sides of nearby subunits. We imagine that by integrating over degrees of freedom, such as atomic coordinates, as capsid proteins fluctuate about their native states, one can arrive at a model in which subunits have excluded volume and asymmetric pairwise bonding interactions between complementary sides.  Several models have been presented in which asymmetric subunits, the capsomers, are represented by conglomerates of spherically symmetric particles with varying interaction strengths \cite{Zhang2004, Chen2005, Rapaport1999, Rapaport2004}. These approaches can describe complex excluded volume shapes.  The approach we take here, however, is simpler, and motivated by the modeling described in Ref. \onlinecite{Schwartz1998}.  Specifically, we use only a single spherical excluded volume per capsomer, and we use internal bond vectors to capture the effects of protein shape and complementarity.

Our model capsomers are of three types: $\bthree$, $\bfour$ and $\bfive$, which contain three, four and five internal capsomer bond vectors, respectively.  These bond vectors, $\mathbf{b}_i^{(\alpha)}$, are pictured in Figure 1.  The index $\alpha$  goes from 1 to $\nb$, where $\nb=3$ for the $\bthree$ model, $\nb=4$ for the $\bfour$ model, and so forth, and the index $i$ goes from 1 to $N$, where $N$ is the number of capsomers in the system.  The vector $\mathbf{r}_i^{(\alpha)} = \mathbf{R}_i + \mathbf{b}_i^{(\alpha)}$  is the position of interaction site $\alpha$   on capsomer $i$, where $\mathbf{R}_i$  is the center of capsomer $i$.  All bond vectors have the same magnitude, $b$.  Within a capsomer frame of reference, the bond vectors are fixed rigidly.  They move only because the capsomer translates and rotates.  This is not to say that proteins do not fluctuate.  Those fluctuations, we imagine, have been averaged over, i.e., integrated out of the model at the level we consider.

\begin{figure}[hbt]
\epsfig{file=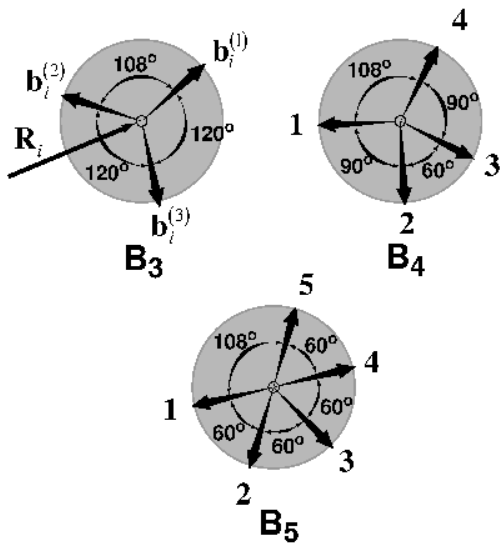,width=.8\linewidth}
\caption{\label{fig:fig1}
Geometry of bond vectors in the $\bthree$, $\bfour$, and $\bfive$ capsomer models.  the center of capsomer $i$ is at $\mathbf{R}_i$.  The angles between indicated bond vectors within a capsomer are specified in degrees.  They do not sum to 360$^{\text{o}}=2\pi$   because the bond vectors are not coplanar.
}
\end{figure}

The net potential energy of interaction among $N$ capsomers, $U(1,2, \ldots, N)$,  is taken to be pair-decomposable, 

\begin{equation}
U(1,2, \ldots, N)=\sum_{i>j=1}^{N}u(i,j)
\label{eq:U}.
\end{equation}
where $u(i,j)$ depends upon the bond vectors and centers of capsomers $i$ and $j$.  The particular form for this pair potential depends upon which of the three models, $\bthree$, $\bfour$ or $\bfive$ is under consideration.  In each case, however, the potential is constructed so that the lowest energy configurations coincide with separate icosohedral clusters of 60 identical capsomers.  These clusters represent capsids with triangulation number (T) of one \cite{Caspar1962}; for example, design $\bfive$ is consistent with Fig.~3 of Ref \onlinecite{Xie1996}.  Our $\bthree$ model is similar to the model considered in Ref.~\onlinecite{Schwartz1998}.

In each model, bond vectors or interaction sites have complementary counterparts.   For example, in the $\bthree$ model, interaction site pairs $(\alpha, \beta) = (1,2)$ and (3,3) are the primary complementary pairs.  This means that a favorable potential energy of interaction between a pair $i$ and $j$ of $\bthree$ capsomers has two ways of occuring: 1) interaction site 1 on one capsomer overlaps with interaction site 2 on the other capsomer, and the respective bond vectors $\mathbf{b}_i^{(1)}$ and $\mathbf{b}_j^{(2)}$ are nearly antiparrallel;  2) interaction site 3 on one capsomer overlaps with interaction site 3 on the other, and $\mathbf{b}_i^{(3)}$ and $\mathbf{b}_j^{(3)}$ are nearly antiparrallel.  The only favorable (i.e., attractive) interactions are those associated with primary complementary pairs.

In addition, there are secondary complementary pairs.  For example, in the $\bthree$ model, with primary complementary pair (1,2), there is the secondary pair (3,3).  This means that a favorable interaction affected by the primary complementary pair (1,2), as described in the previous paragraph, also requires that $\mathbf{b}_i^{(3)}$ and $\mathbf{b}_j^{(3)}$ are nearly coplanar.  Similarly, for the primary complementary pair (3,3), the secondary pair is either (1,2) or (2,1), meaning that if $\mathbf{b}_i^{(3)}$    and $\mathbf{b}_j^{(3)}$  are antiparallel, favorable interactions result only if $\mathbf{b}_i^{(1)}$ and $\mathbf{b}_j^{(2)}$ are nearly co-planar and $\mathbf{b}_i^{(2)}$ and $\mathbf{b}_j^{(1)}$ are nearly co-planar.  Of course, because the capsomers are rigid bodies, $\mathbf{b}_i^{(1)}$ and $\mathbf{b}_j^{(2)}$ being co-planar implies $\mathbf{b}_i^{(2)}$ and $\mathbf{b}_j^{(1)}$ are co-planar.  The primary and secondary pairs for each of the models are listed in the entrees to Table I.  Local bonding associated with these complementarities and resulting capsid structures are illustrated in Figure 2.  In creating these pictures, it is imagined that excluded volume interactions prohibit an interaction site from participating simultaneously in more than one favorable complementary interaction, as is the case for the models we describe.

The dependence of subunit-subunit interactions on the orientation of primary and secondary pairs incorporates the fact that there is a driving force for subunits to align complementary regions to maximize the contact between complementary residues.  Capsid curvature in the minimum energy orientation arises from that fact that the angles between bond vectors on a given subunit do not sum to $2\pi$.

\begin{table}[hbt]
\begin{tabular*}{.8\linewidth} {@{\extracolsep{\fill}}cccc} 
  \multicolumn{1}{c}{}&\multicolumn{1}{r}{\large{Primary}}&\multicolumn{2}{c}{\large{Secondary}} \\ \cline{2-4}
   &$\alpha\quad\beta$&$\gamma\quad\epsilon$&$\eta\quad\nu$\\ \cline{2-4}
\large{B3} & \begin{tabular}{c} $1\quad2$ \\ $2\quad1$ \\ $3\quad3$ \\ \end{tabular}& 
\begin{tabular}{c} $2\quad1$ \\ $1\quad2$ \\ $1\quad2$ \\ \end{tabular}& 
\begin{tabular}{c} $3\quad3$ \\ $1\quad2$ \\ $2\quad1$ \\ \end{tabular} \\
 & & & \\
\large{B4} & \begin{tabular}{c} $1\quad4$ \\ $2\quad3$ \\ $3\quad2$ \\ $4\quad1$ \\ \end{tabular}& 
 \begin{tabular}{c} $2\quad3$ \\ $1\quad4$ \\ $4\quad1$ \\ $3\quad2$ \\ \end{tabular}&\\ 
&&& \\ 

\large{B5} & \begin{tabular}{c} $1\quad5$ \\ $2\quad2$ \\ $3\quad4$ \\ $4\quad3$ \\ $5\quad1$ \\ \end{tabular}& 
\begin{tabular}{c} $5\quad1$ \\ $3\quad1$ \\ $2\quad5$ \\ $5\quad2$ \\ $1\quad5$ \\ \end{tabular}& 
\begin{tabular}{c} $2\quad4$ \\ $1\quad3$ \\ $4\quad3$ \\ $3\quad4$ \\ $4\quad2$ \\ \end{tabular} \\
\end{tabular*}
\begin{eqnarray*}
  \cos\left(\theta_{ij}^{(\alpha\beta)}\right)=-\mathbf{b}_i^{(\alpha)}\cdot \mathbf{b}_i^{(\beta)}/b^2 \quad \quad \mathbf{R}_{ij}=\ri-\rj 
\end{eqnarray*}
\begin{eqnarray*}
\cos\left(\phi_{ij}^{(\alpha\beta,1)}\right)&=&\frac{\left(\mathbf{b}_i^{(\gamma)}\times \mathbf{R}_{ij}\right) \cdot \left(\mathbf{R}_{ij} \times \mathbf{b}_i^{(\epsilon)}\right)}{\left|(\mathbf{b}_i^{(\gamma)}\times \mathbf{R}_{ij}\right| \left|\mathbf{R}_{ij} \times \mathbf{b}_i^{(\epsilon)}\right|} \\ 
\cos\left(\phi_{ij}^{(\alpha\beta,2)}\right)&=&\frac{\left(\mathbf{b}_i^{(\eta)}\times \mathbf{R}_{ij}\right) \cdot \left(\mathbf{R}_{ij} \times \mathbf{b}_i^{(\nu)}\right)}{\left|(\mathbf{b}_i^{(\eta)}\times \mathbf{R}_{ij}\right| \left|\mathbf{R}_{ij} \times \mathbf{b}_i^{(\nu)}\right|}
\end{eqnarray*}
\caption{\label{tab:tab1}
Primary and secondary complementary pairs and associated angles for the three capsomer models.
}
\end{table}

\renewcommand{\arraystretch}{1}

{\bf Pair potential.}  
The potential energy of interaction between two capsomers, say 1 and 2, is taken to have a spherically symmetric repulsive part, $u_0{\left(|\mathbf{R}_2-\mathbf{R}_1|\right)}$, and an attractive part that depends upon both $\mathbf{R}_2-\mathbf{R}_1$ and the bond vectors associated with the two capsomers, 
\begin{eqnarray}
u(1,2)&=& u_0(|\mathbf{R}_2-\mathbf{R}_1|) \nonumber \\ 
& & +u_1\left(\mathbf{R}_2-\mathbf{R}_1,\left\{\mathbf{b}_2^{(\alpha)}\right\},\left\{\mathbf{b}_1^{(\gamma)}\right\}\right)
\label{eq:uij}.
\end{eqnarray}
For the repulsion, we have chosen the Weeks-Chandler-Andersen \cite{Andersen1972} potential,
\begin{figure}[hbt]
\epsfig{file=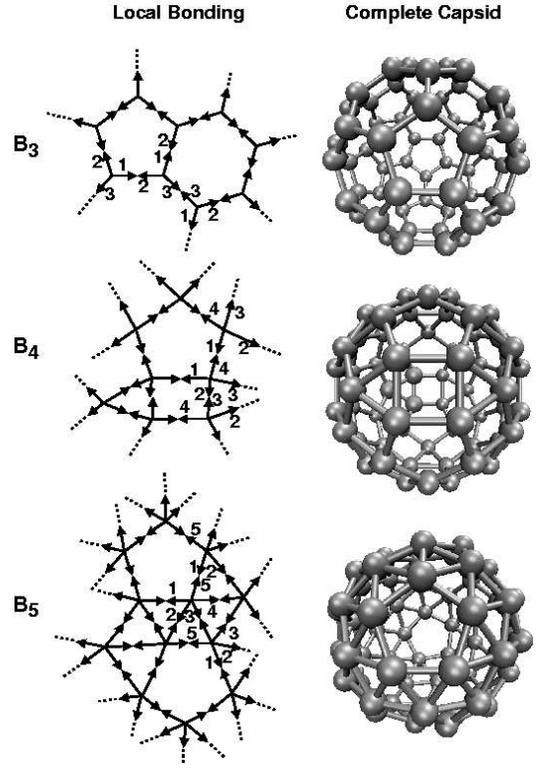,width=.8\linewidth}
\caption{\label{fig:capsiddesigns}
Complementary pairs and bonding of capsomers.  The first column specifies the model, the second illustrates the local bonding consistent with the complementary pairs of bond vectors, and the third illustrates the resulting complete capsid, with bonds depicting the attractive interactions resulting from complementary pairs.  The pictures of complete capsids and all simulation snapshots shown in this work were generated in VMD \cite{Humphrey1996}. The size of the spheres in these pictures has been reduced to aid visibility; parameters are chosen such that the minimum energy distance between neighboring capsomers is at the minimum in the WCA potential, Eq.~\ref{eq:wca}.}
\end{figure}
\begin{eqnarray}
u_0(R) &=& 4\epsilon\left[(\sigma/R)^{12} - (\sigma/R)^{6} + 1/4 \right], \quad R<2^{1/6}\sigma \nonumber \\
&=& 0, \quad R\ge 2^{1/6} \sigma  
\label{eq:wca}.
\end{eqnarray}
For the attractions we have chosen
\begin{eqnarray}
u_1\left(\mathbf{R}_2-\mathbf{R}_1,\left\{\mathbf{b}_2^{(\sigma)}\right\},\left\{\mathbf{b}_2^{(\gamma)}\right\}\right) \qquad \qquad \nonumber \\ = {\displaystyle \sum}_{\alpha\beta}^{'} u_{\text{att}}\left(\left|\mathbf{r}_2^{(\sigma)}-\mathbf{r}_1^{(\beta)}\right|\right)s_{\alpha\beta}(1,2)
\label{eq:Ub},
\end{eqnarray}  
where the primed sum is over primary complementary pairs,
\begin{eqnarray}
u_{\text{att}}(r) &=& 4\eb \Bigg[\left(\frac{\sigma}{r+2^{1/6}\sigma}\right)^{12}-\left(\frac{\sigma}{r+2^{1/6}\sigma}\right)^{6}
\nonumber \\
 & & \quad -\left(\frac{\sigma}{\rc}\right)^{12}+\left(\frac{\sigma}{\rc}\right)^{6}\Bigg], \quad r+2^{1/6} \sigma<\rc \nonumber \\
&=& 0, \quad r+2^{1/6} \sigma\ge\rc
\label{eq:eb},
\end{eqnarray}
which has its minimum value, $-\eb$, when the separation of complementary pair interaction sites is zero, and $s_{\alpha\beta}(1,2)$ is the switching function, given by
\begin{eqnarray}
s_{\alpha\beta}(1,2)&=&\frac{1}{8}\left[\cos\left(\pi\theta_{12}^{(\alpha\beta)}/\tm\right)+1\right] \nonumber \\ 
& & \times\left[\cos\left(\pi\phi_{12}^{(\alpha\beta,1)}/\pmax\right) +1\right]\nonumber \\ 
& & \times \left[\cos\left(\pi\phi_{12}^{(\alpha\beta,2)}/\pmax\right) +1\right]
\label{eq:sab}
\end{eqnarray}
for models $\bthree$ and $\bfive$, and by
\begin{eqnarray}
s_{\alpha\beta}(1,2)&=&\frac{1}{4}\left[\cos\left(\pi\theta_{12}^{(\alpha\beta)}/\tm\right)+1\right] \nonumber \\ 
& &\times \left[\cos\left(\pi\phi_{12}^{(\alpha\beta,1)}/\pmax\right) +1\right]
\label{eq:sabb}
\end{eqnarray} 
for model $\bfour$.  The angle variables used in these expressions are defined in Table I.  Notice from that table, specifying a specific primary pair of complementary bonds prescribes specific corresponding secondary pairs.  The switching function goes smoothly from 1 to 0 as the angle variables  $\theta_{12}^{(\alpha \gamma)}$, $\phi_{12}^{(\alpha \gamma,1)}$ and $\phi_{12}^{(\alpha \gamma,2)}$ change from 0 to $\tm$ , $\pmax$ and $\pmax$, respectively.  Increase of these maximum angles $\tm$ and $\pmax$ increases the configuration space in which two nearby subunits attract each other, but also weakens the driving force toward the minimum energy orientation.

{\bf Dynamical simulations}  
Dynamical trajectories were calculated using Brownian dynamics, in which particle motions are calculated from Newton's laws with forces and torques arising from subunit-subunit interactions as well as drag and a random buffeting force due to the implicit solvent.  We use the following coupled equations of motion 
\begin{eqnarray}
\dot{\ri} & = & \gamma\f_i+\df_i \nonumber \\
\oomega_i & = & \gamr \ttau_i+\dtau_i
\label{eq:eom}
\end{eqnarray}   
where $\oomega$ is the angular velocity, the force is given by 
\begin{equation}
\f_i=-\partial U/ \partial \ri
\label{eq:force},
\end{equation}   
and the torque is given by
\begin{equation}
\ttau_i=-\sum_{\alpha}\bbond_i^{(\alpha)} \times \left(\partial U/ \partial \bbond_i^{(\alpha)} \right) 
\label{eq:torque}
\end{equation}   
while $\df_i$ and $\dtau_i$ are a random force and torque, with covariances given by 
\begin{eqnarray}
\left\langle \df_i(t)\df_j(t')\right\rangle & = & \mathbf{1} \mathrm{\delta}(t-t')\mathrm{\delta}_{ij} 2 \kt/\gamma \nonumber \\
\left\langle \dtau_i(t)\dtau_j(t')\right\rangle & = & \mathbf{1} \mathrm{\delta}(t-t')\mathrm{\delta}_{ij} 2 \kt/\gamr 
\label{eq:covariance}
\end{eqnarray}   
where $\mathbf{1}$ is the identity matrix.
The friction coefficients for translation and rotation are $\gamma$ and $\gamr$, respectively, and $\kt$ is the thermal energy.

\begin{table}[hbt] 
{\footnotesize
\begin{tabular} {|c|c|l|}       \hline
Parameter & Value & Definition\\ \hline
$\epsilon/\kt$ & 1  &  WCA energy parameter, Eq. \ref{eq:wca}\\
$\eb/\kt$ & $9 - 22$  &  Attractive energy strength, Eq. \ref{eq:eb}\\
$b/\sigma$ & $2^{-5/6}$  &  Bond vector length\\
$\gamr/\gamma \sigma^2$ & 0.4   &  Rotational friction coefficient, Eq. \ref{eq:eom}\\
$\pmax$ (rad) & 3.14  &  maximum dihedral angle, Eq. \ref{eq:sab}\\
$\tm$ (rad) & $0.1-3.0$ &  maximum bond angle, Eq. \ref{eq:sab}\\
$L/\sigma$ & $11-100$ &  Simulation boxsize\\
$N$ & 1000 &    Number of subunits\\
$\conc=N \sigma^3L^{-3}$ & $0.001-0.75$  &   Concentration of subunits\\
$r_{\text{c}}/\sigma$ & 2.5  &  Attractive energy cutoff distance\\
$\delta t/t_0$ & 0.006   &  Timestep\\
$\tobs/t_0$ & $\tf$  & Final observation time, $10^8$ steps\\
\hline
\end{tabular}
}
\caption{\label{tab:parameters}
Parameter values used for dynamical simulations in this work, where $\sigma$ is the unit of length, $\kt$ is the thermal energy, $\gamma$ is the translational friction constant, Eq.~\ref{eq:eom}, and $t_0\equiv \gamma \sigma^2/(48\kt)$ is the unit time. 
 }
\end{table}

In our implementations of these equations, rigid body rotations were performed with quaternions \cite{Allen1987} and rotational and translational displacements were calculated using the second order stochastic Runge-Kutta method \cite{Branka1999, Heyes2000}, as described in Appendix A. Periodic boundary conditions were used to simulate a bulk system. We employed reduced units in which the particle diameter $\sigma=1$, $\kt$ is the unit of energy, and time is scaled by $t_0\equiv \gamma \sigma^2/(48\kt)$.  Each trajectory considered $N=1000$ subunits and ran for $10^8$ steps, usually with a time step of $0.006$. The values of all parameters used in this work are documented in Table II. If units of length, $\sigma$, and temperature, $T$, are chosen to be $\sigma=2$ nm and $T=300$ K, the final observation time after $10^8$ steps is $\tobs=227.5$ $\mathrm{\mu}$s, subunit concentrations, $\conc$ range from $2.08 \times 10^{-4} - 0.156$ mol/L, and binding energies, $\eb$, range from $5.4-13.2$ kcal/mol.

\begin{figure}[hbt]
\epsfig{file=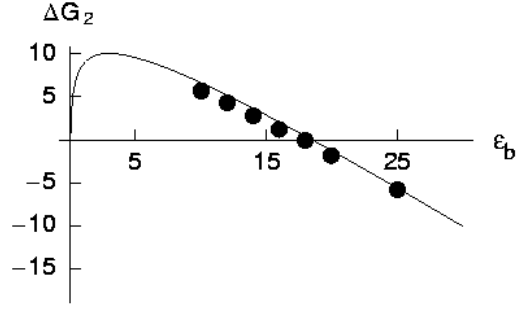, width=.8\linewidth}
\caption{\label{fig:bindingfreeenergy}
Binding free energies for dimerization  calculated from Eqs.~\ref{eq:sdimer},\ref{eq:saddlepoint} (line) and Monte Carlo simulations (points).  Free energies are with reference to a standard state with volume fraction of 1 and free rotational motion, and the maximum angle parameters, defined in Eq.~\ref{eq:sab}, are $\tm=\pmax=0.5$.   
}
\end{figure}

\begin{figure}[hbt]
\epsfig{file=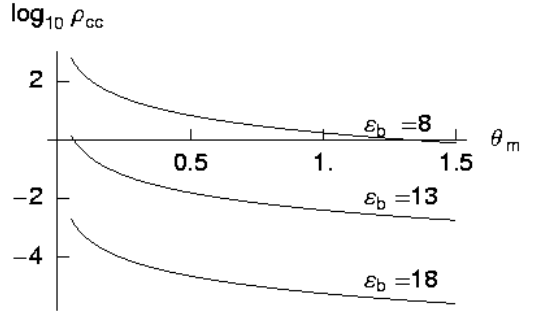, width=.8\linewidth}
\caption{\label{fig:ccc}
The thermodynamic critical subunit concentration for capsid formation, $\ccc$, as calculated from Eqs~\ref{eq:saddlepoint}, \ref{eq:ccc}, and \ref{eq:fnc} for design $\bthree$ and $\pmax=\pi$.  Above these subunit concentrations, most subunits will be found in complete capsids at equilibrium.  
}
\end{figure}

\begin{figure*}[hbt]
\epsfig{file=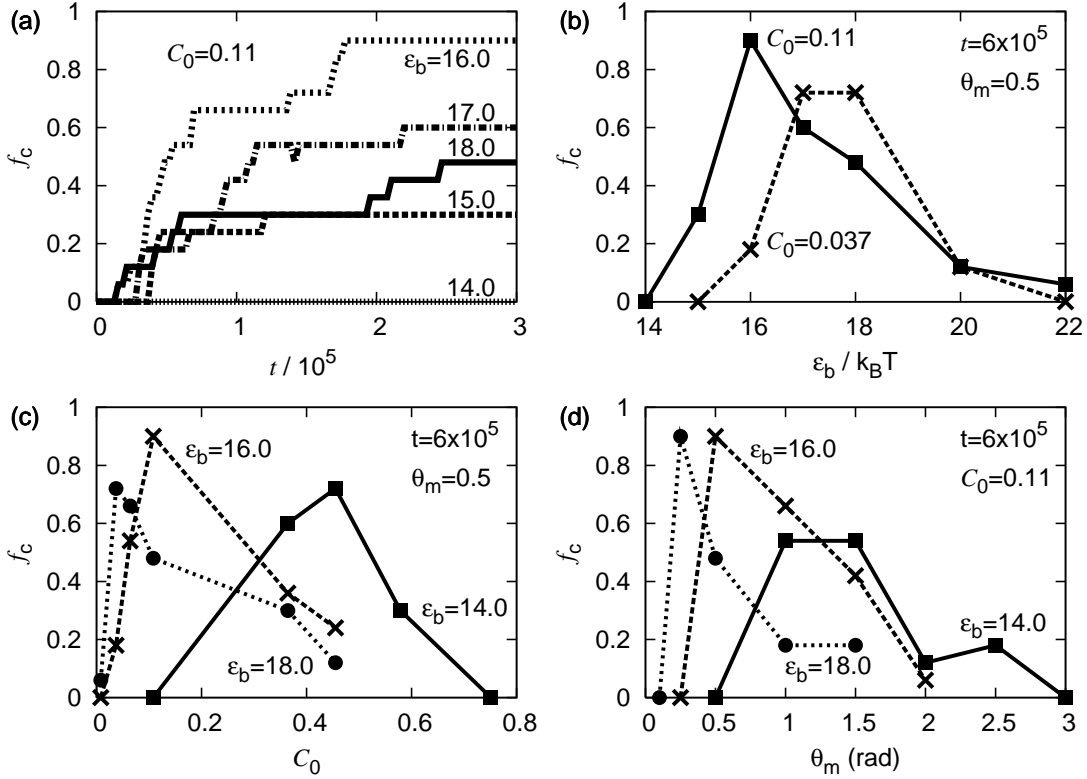,width=.8\textwidth}
\caption{\label{fig:fc}
Examples of the influence of system parameters on assembly dynamics for design $\bthree$. (a) The fraction of complete capsids versus time, $\ct$, is shown for $\tm=0.5$ and $\conc=N\sigma^3/L^3=0.11$ at varying $\eb$ illustrating the sigmoidal shape of capsid yields.  Note that variations of $\ct$ are in discrete units of 0.06 because there are 1000 subunits and each complete capsid has 60 subunits.   Variation of the final mass fraction of complete capsids, $\ct$, is shown in (b)-(d):  (b) variation with $\conc=0.11$ and $\tm=0.5$, (c) variation with $\conc$ at several values of $\eb$ with $\tm=0.5$, and (d) variation with $\tm$ for several values of $\eb$ and $\conc=0.11$.  Note that $\eb$ does not denote the free energy to bind; there is a significant entropy penalty, calculated in Eq.~\ref{eq:saddlepoint}. 
}
\end{figure*}

\section{Thermodynamics of capsid assembly}
\label{sec:thermo}
The equilibrium concentrations of free subunits (monomers) and capsid intermediates can be related by the law of mass action \cite{Chandler1987}
\begin{equation}
\rho_n \sigma^3=\left(\rho_1 \sigma^3\right)^n \exp(-\beta \Delta G_n)
\label{eq:massaction}
\end{equation}
where $\rho_n$ is the number density of an intermediate with $n$ subunits, $\sigma$ is the molecular dimension, $\beta=1/k_{\text{B}}T$ is the inverse of the thermal energy, and $\Delta G_n$ is the driving force to form an intermediate of size $n$.  The driving force for assembly comes from the fact that subunits experience a favorable energy, $\eb$, upon binding, but subunits also face an entropic penalty, which depends on the number of bonds and the local bonding network.  

The free energy for making a single bond to form a dimer can be determined by calculating the ratio of the partition functions for two bound subunits and two free subunits \cite{Erickson1981,Ben-Tal2000}
\begin{eqnarray}
q_2/q_1^2 &=& \frac{1}{8\pi^2} \int d \mathbf{R}_2  \int d \bm{\Omega}_2   \\ 
& & \exp\left[-\beta u(\mathbf{R}_1,\mathbf{R}_2,\{\mathbf{b}_1\},\{\mathbf{b}_2\})\right]H(1,2)\nonumber
\label{eq:q}
\end{eqnarray}
where $u$ is defined in Eq.~\ref{eq:uij}, $\bm{\Omega}_2$ describes the Euler angles of subunit 2, which specify the set of bond vector orientations, $\{\mathbf{b}_2\}$, and $H(1,2)$ is unity when $ u(\mathbf{R}_1,\mathbf{R}_2,\{\mathbf{b}_1\},\{\mathbf{b}_2\})< -2 k_{\text{B}}T$, and zero otherwise. In other words, we define two capsomers as bound if their potential energy of interaction is lower than $-2\kt$.  The free subunits are taken to be at a standard state with unit density and free rotation, and the coordinate system is centered on $\mathbf{R}_1$.

  Expansion of $u(1,2)$ to quadratic order in each coordinate about the minimum in the potential gives
\begin{equation}
\Delta G_2 = -\kt \ln q_2/q_1^2 =  -\eb - T \sbond 
\label{eq:sdimer}
\end{equation}
with 
\begin{equation}
\sbond/k_{\text{B}} \approx -
\frac{3}{2} \ln \left. \frac{2 \pi \beta \partial^2 u_{\text{att}}(r)}{\partial r^2}\right|_{r=0} + \frac{1}{2} \ln \frac{2 \beta \eb^{3} \pi^{7}}{\tm^4\pmax^2} 
\label{eq:saddlepoint}
\end{equation}  
where the two terms represent translational and rotational entropy, respectively.
This result is compared to binding free energies calculated with Monte Carlo simulations in Fig.~\ref{fig:bindingfreeenergy}. 

\begin{figure*}[hbt]
\epsfig{file=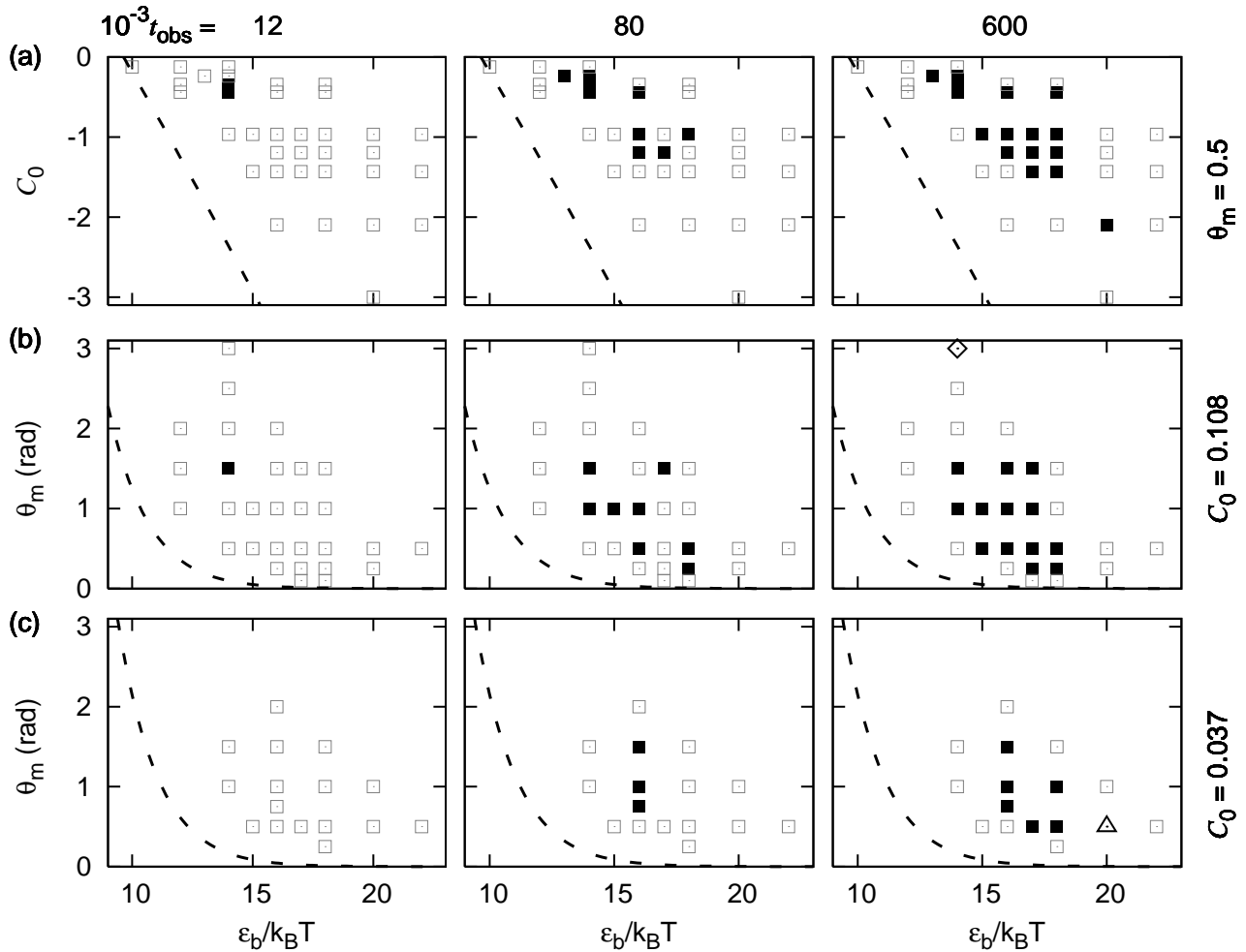,width=\textwidth-1cm}
\caption{\label{fig:kpd}
Changing model parameters reveals the kinetic phase diagram for design $\bthree$.  Filled points denote parameter values  for which $30\%$ of subunits are in complete capsids ($\ct\ge 30\%$) by the observation time,  $\tobs$, open points indicate parameter values for which $\ct<30\%$, and the dashed lines indicate the location of the thermodynamic critical surface, calculated with Eq.~\ref{eq:ccc}. The first, second, and third columns correspond to observation time $\tobs=1.2\times10^4$, $8\times10^4$, and $6\times10^5$, respectively.  The top row (a) shows cross-sections through $\conc$ and $\eb$ with $\tm=0.5$. The second and third row show cross-sections through $\tm$ and $\eb$, with (b)  $\conc = 0.11$ and (c) $\conc=0.037$.  In each case the region covered by filled points roughly defines a cross-section of parameter space within which assembly is kinetically possible. Simulation snapshots corresponding to the $\diamond$ in the right hand panel in row (b) and the $\triangle$ in the right hand panel in row (c) are shown in Fig. \ref{fig:kt}. 
}
\end{figure*}

While there are many possible capsid structures consistent with most larger values of $n$, there is only one structure consistent with a complete capsid, which has $n=N_{\text{c}}$ subunits ($N_{\text{c}}=60$ for the capsids studied in this work).  The fact that misformed capsids and intermediates are generally not observed implies that $\Delta G_n$ is sharply peaked at $n=N_{\text{c}}$; defects that lead to larger or smaller capsids are unfavorable. There is a threshold density, $\ccc$, at which the fraction of subunits in capsids becomes significant \cite{Ben-Shaul1994,Bruinsma2003,Maibaum2004,Kegel2004}
\begin{equation}
\ln \ccc a^3 \approx \beta \Delta G_{N_{\text{c}}}/N_{\text{c}}
\label{eq:ccc}.
\end{equation}
By analogy with Eq.~\ref{eq:sdimer}, the free energy of a complete capsid can be written as
\begin{equation}
\Delta G_{N_c} = - N_{\text{c}} \nb \eb/2 - T (N_c-1) \sbond(\nb) 
\label{eq:fnc},
\end{equation}
where $\sbond(\nb)$ is the entropy penalty for a subunit in a complete capsid, where each subunit has $\nb$ bonds.
If we neglect the dependence of the entropy penalty on the number of bonds [i.e. assume $\sbond(\nb)\approx \sbond$], we can use Eqs.~\ref{eq:saddlepoint}, \ref{eq:ccc}, and \ref{eq:fnc} to calculate $\ccc$.  Values of $\ccc$ for capsid design $\bthree$ ($\nb=3$) with $\pmax=\pi$ (the value used for all dynamical simulations with this work) are shown in Fig. \ref{fig:ccc}, and are compared to kinetic assembly results in Fig.~\ref{fig:kpd}.

\section{Kinetics of Capsid Assembly}
\label{sec:results}
{\bf Capsid formation rate curves are sigmoidal.}
We have considered capsid assembly dynamics for design $\bthree$ (see Figs.~1 and 2) over ranges of subunit concentrations, $\conc$ (reported in dimensionless units, $\conc=N\sigma^3/L^3$), binding energies, $\eb$, and maximum binding angles $\tm$.    The results we present use $\pmax=\pi$; the effect of varying $\pmax$ is similar to, but less dramatic than that of varying $\tm$.  Dynamics of different capsid designs are discussed below.

The fraction of subunits in completed capsids, $\fc$, is shown as a function of time for several binding energies in Fig.~\ref{fig:fc}a. In all cases for which significant assembly occured, the rate of capsid formation has a roughly sigmoidal shape. This is a general feature of assembly reactions \cite{Endres2002} that can be understood as follows.  There is an initial lag phase during which capsid intermediates form and progress through the assembly cascade, followed by a rapid growth phase during which these intermediates assemble into complete capsids. Finally, growth slows when monomers (free subunits) are depleted and the remaining capsid intermediates are unable to bind with each other. 
\begin{figure*}[hbt]
\epsfig{file=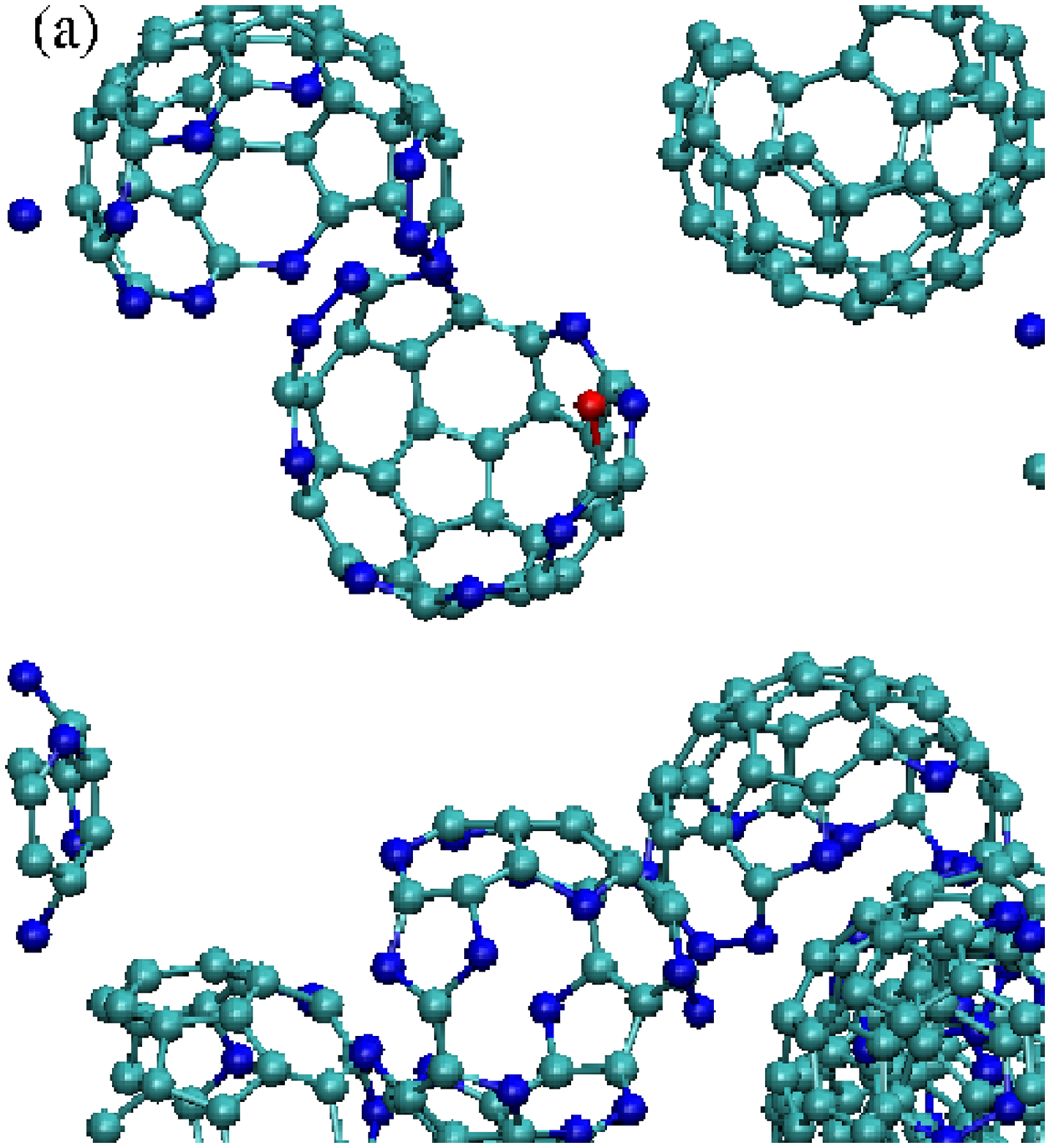,height=\textheight/3}
\hspace{1cm}
\epsfig{file=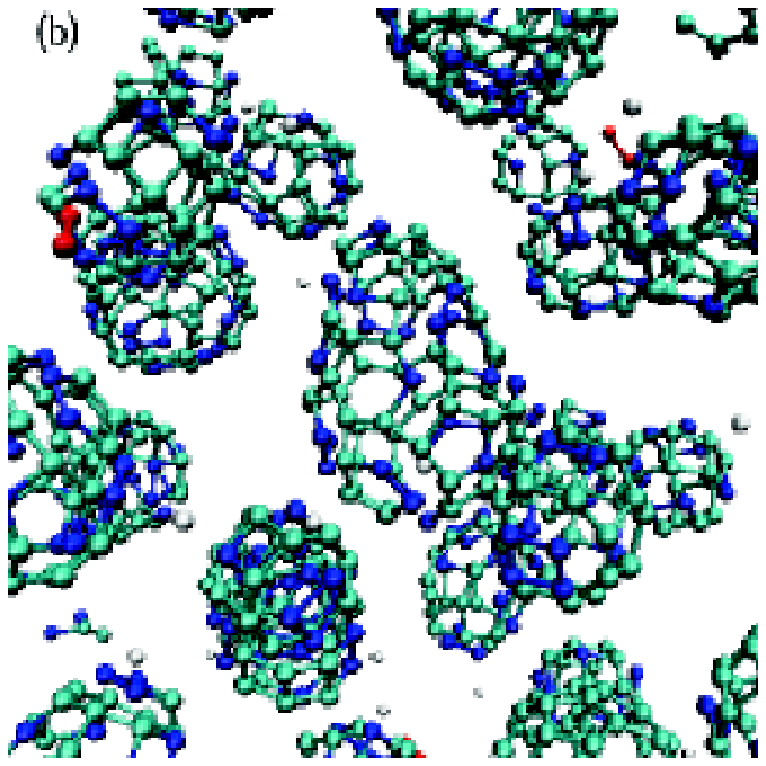,height=\textheight/3}
\caption{\label{fig:kt}
Snapshots corresponding to unassembled points in Fig. \ref{fig:kpd}, which illustrate the two kinetic traps described in the text.  (a) Parameter values corresponding to the $\diamond$ in Fig. \ref{fig:kpd} allowed rapid capsid initiation and growth, which depleted monomers and small intermediates before capsids were completed. (b) At parameter values corresponding to the $\triangle$, strained bonds are incorporated into growing capsids.  These snapshots have zoomed in on a representative portion of the system to aid visibility.  Subunit colors indicate the number of bonds: white = 0, red = 1, turquoise = 2, and dark blue = 3 bonds.
}
\end{figure*}

{\bf Final capsid yields are non-monotonic with respect to parameter values, but high yields are possible.}  The fraction of subunits in complete capsids, $\fc$, at the final observation time, $t=\tf$, is shown in Fig \ref{fig:fc}.  As $\conc$, $\eb$, or $\tm$ increase, intermediates form and grow more rapidly, and thus capsid yields increase to as high as 90\%, meaning 15 of the 16 possible capsids were completed.  One of the primary results of this study is that a particle model that does not include details such as heterogeneous nucleation or conformational changes can predict such high capsid yields. 

Although capsids form more quickly as parameter values are increased, saturation of growth also occurs sooner and capsid yields are non-montonic in each parameter.  The sensitivity of capsid yields to parameters seen in Fig. \ref{fig:fc} is further illustrated with a kinetic phase diagram in Fig. \ref{fig:kpd}.  It demonstrates the coupled dependencies of capsid yields on system parameters.  Phases are partitioned according to whether or not there is significant assembly, arbitrarily chosen as $\ct \ge 30\%$. 

{\bf The non-monotronic variation of capsid yields with parameter values arises due to competition between faster capsid growth and kinetic traps.}  
The initial steps in the assembly cascade result in the formation of fewer bonds than later steps.  If the attractive energy of these bonds is not sufficient to overcome entropic loss the initial steps are uphill on a free energy barrier and hence, are slow.  For parameter sets that are near $\ccc$ (see Fig. \ref{fig:ccc}), where half of the subunits are in complete capsids at equilibrium, the fact that a complete capsid has many more bonds than initial assembly products indicates that the free energy barrier must be many times the thermal energy, $\kt$.  Significant assembly, therefore, does not occur within the finite assembly times we consider until parameter values are much higher than the thermodynamic critical values, and we identify a kinetic lower critical surface (LCS) in Fig.~\ref{fig:kpd} that bounds the regions with significant assembly from below and to the left.  We consider results at finite observation times because capsid assembly reactions are limited in-vivo by proteolysis times and in-vitro by experimental observation times. 

  Increasing parameters increases the overall rate of capsid growth: higher subunit concentrations, $\conc$, result in more frequent subunit collisions, higher values of $\tm$ increase the likelihood of binding upon a collision, and higher values of the binding energy, $\eb$, decrease the rate of the reverse reaction (subunit unbinding).  As parameter values cross the LCS, faster capsid growth leads to significant capsid yields, as seen in Fig.~\ref{fig:fc}.  At even higher values, however, assembly becomes frustrated by two kinetic traps (see Fig. \ref{fig:kt}) and we identify an upper critical surface (UCS) in Fig \ref{fig:kpd} to the top and right of the regions in which assembly is kinetically accessable.  Because of these kinetic traps, assembly only occurs at subunit-subunit binding energies that are much smaller than values calculated from atomistic potentials in Ref. \onlinecite{Reddy1998} (see Table 1 of that Ref.). When the binding entropy (see Eq.~\ref{eq:saddlepoint}) is included, however, the resulting free energies are consistent with association constants fit to assembly experiments with Hepatitis B capsids in Ref.~\cite{Ceres2002}.  

We present results at three observation times to show how the distance between assembly boundaries expands in all directions as time increases.  The rough boundary of the kinetically accessable region in Fig.~\ref{fig:kpd} is a measure of the statistical uncertainty that results because each data point describes a single stochastic trajectory.  For trajectories run with different random number seeds at a given set of parameter values, the final number of complete capsids typically did not vary by more than one capsid; however, variations during the rapid growth phase were larger.

The kinetic trap depicted in Fig \ref{fig:kt}a arises when progress through initial assembly steps is too rapid, allowing so many capsids to initiate that the pool of monomers and small intermediates becomes depleted before a significant number of capsids are completed. If the remaining partial capsids have non-complementary geometries, further binding can only proceed upon disassembly. This kinetic trap has been seen in experiments \cite{Zlotnick2000} and predicted theoretically \cite{Zlotnick1994}; this theory, however, assumes that only monomers can add to partial capsids. As discussed below, binding of capsid intermediates is an important mode of assembly and growth does not become frustrated until only intermediates with non-complementary geometries remain.

The kinetic trap just described may not limit assembly {\em in vivo}, where there is a continual supply of new capsid proteins. Subunit bonding in configurations not consistent with a complete capsid (misbonding), however, can lead to a kinetic trap that could frustrate assembly even with an unlimited supply of subunits. It is not surprising that misbonding occurs more frequently as $\tm$ increases, since there is a smaller driving force toward the minimum energy orientation. As noted by Berger and co-workers \cite{Schwartz1998}, though, increasing concentration and binding energy can stabilize subunits with strained bonds, and the higher rate of capsid growth under these conditions can cause misbonded subunits to become trapped in a growing capsid by further addition of subunits. Because so many assembly pathways that do not lead to complete capsids are available at these parameter values, the minimum energy configuration, with complete capsids, is seldom realized and misformed capsids with spiraling or multi-shelled configurations dominate, as shown in Fig.~\ref{fig:kt}b. Progression from this state to a completed capsid is extremely slow because breakage of many bonds is required. It would be difficult to assess the importance of configurations such as that shown in Fig.~\ref{fig:kt}b with models that have assumed assembly pathways.  

\begin{figure}[hbt]
\epsfig{file=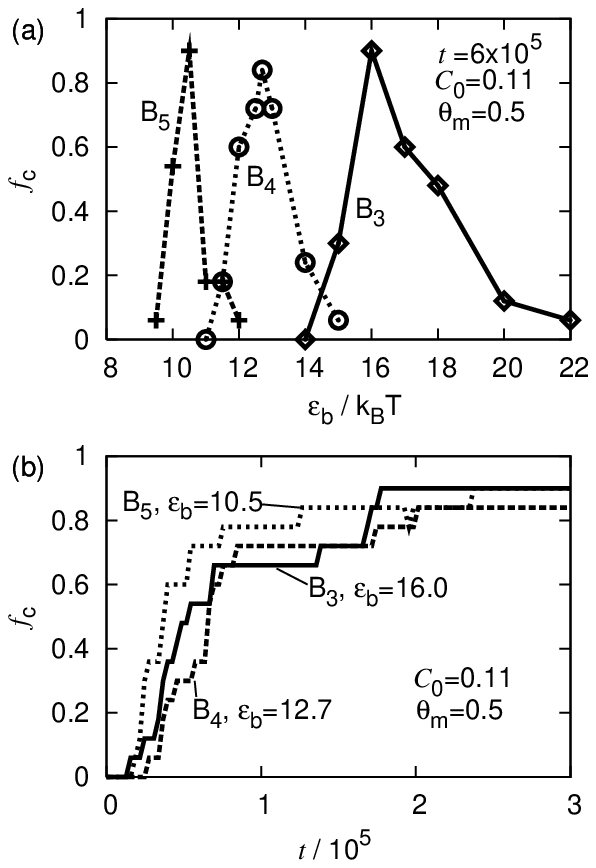,width=.8\linewidth}
\caption{\label{fig:optass}
Assembly kinetics are similar for each capsid design, $\bthree$, $\bfour$, and $\bfive$.  (a) The final capsid yields, $\ct$ at $\tobs=\tf$, are shown versus binding energy for each capsid design for $\conc=0.11$ and $\tm=0.5$.  (b) The assembly time series are shown for the optimal values of $\eb$ from (a) ($\eb=16.0$ for $\bthree$, 12.7 for $\bfour$, and 10.5 for $\bfive$).
}
\end{figure}

{\bf $\bfive$ capsids grow primarily through additions of individual subunits, while combination of clusters is essential for assembly of $\bfour$ and $\bthree$ capsids.}  
The variations of final capsid yields with $\eb$ for the capsid designs $\bthree$, $\bfour$, and $\bfive$ (see Fig \ref{fig:capsiddesigns}) are shown in Fig \ref{fig:optass}a.  Although assembly occurs within different ranges of $\eb$ for each design, assembly kinetics within these ranges are similar, as shown in Fig \ref{fig:optass}b.  In addition, the optimal assembly ($\ct\approx 0.9$) for each design occurs at approximately the same value of $\nb \eb \approx 50$, meaning that the complete capsids all have about the same stability.  Variation of assembly with $\conc$ and $\tm$ (not shown) is also similar for each design.
\begin{figure}[hbt]
\epsfig{file=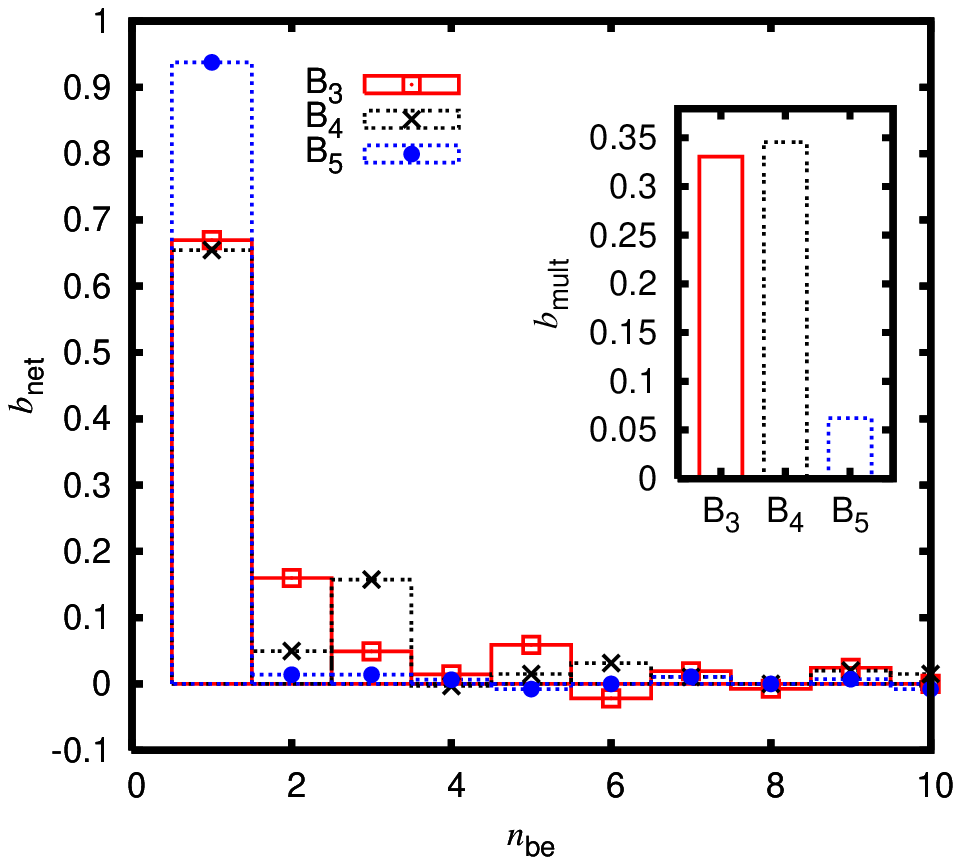, width=.8\linewidth}
\caption{\label{fig:nbe}
$\bfive$ capsids grow primarily through additions of individual subunits, while binding of multimers to growing capsids is essential for assembly of $\bfour$ and $\bthree$ capsids.  The fraction of binding, $\nfb$ (see Appendix B), for each cluster sizes $\nbe$ is shown for each design, and the total binding contribution for multimers, $\bmult=\sum_{\nbe>1}\nfb(\nbe)$, is shown in the inset.  Statistics were measured for the parameters used in Fig. \ref{fig:optass} through $t=2.25\times 10^5$, by which time the majority of assembly was completed for these parameter values.
}
\end{figure}

Although the capacity to assemble spontaneously is similar for each capsid design, the mechanism of assembly (near optimal assembly) for $\bfive$ is qualitatively different from that for $\bfour$ and $\bthree$.  We evaluated assembly mechanisms from simulations by tabulating the size, $\nbe$, of the smallest intermediate involved in each binding or unbinding event (see Appendix B).  The net contribution to capsid growth by intermediates of size $\nbe$, $\nfb(\nbe)$, is shown for each capsid design in Fig \ref{fig:nbe}.  While about 33\% of subunits assembled as multimers ($\nbe>1$) for $\bthree$ and $\bfour$ capsids, multimer binding accounted for only 6\% of all binding for $\bfive$ capsids.
\begin{figure}[hbt]
\epsfig{file=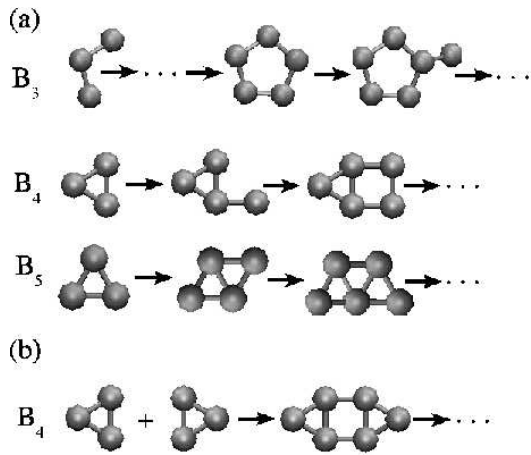,angle=270,width=.8\linewidth}
\caption{\label{fig:asspath}
(a) Examples of the assembly pathways that are available if only monomers can add to growing capsids.  Once a $\bfive$ dimer is formed, subsequent monomers can always form two or more bonds, whereas all assembly paths for $\bthree$ and $\bfour$ require formation of single bonds.  (b) One example of a multimer binding step for design $\bfour$ that avoids formation of single bonds.
}
\end{figure}

The influence of capsid design on mechanism can be understood by examining assembly pathways that are available if growth occurs only through monomer additions, such as those shown in Fig.~\ref{fig:asspath}. Once dimerization occurs for $\bfive$, all subsequent monomers can add in such a way that two or more bonds are formed. For parameters values at which optimal assembly occurs, the formation of a single bond is unfavorable due to entropy loss, but the formation of two or more bonds is favorable. Consequently, the approximate projection of free energy onto cluster size (see Appendix C) shown in Fig. \ref{fig:fenergy} is monotonically decreasing after formation of a dimer.  
\begin{figure}[hbt]
\epsfig{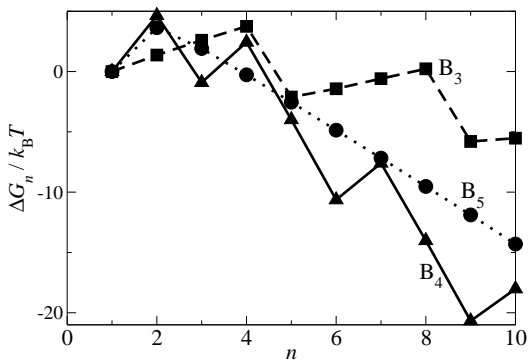} 
\caption{\label{fig:fenergy}
Free energy projections onto the number of particles in a capsid, $n$, for each design, relative to monomers in solution at $\conc=0.11$ with $\tm=0.5$.  The values of $\eb$ are the optimal values from Fig.~\ref{fig:optass}, 16.0, 10.5, and 12.7 for the $\bthree$, $\bfour$, and $\bfive$ designs, respectively.  Free energies were obtained with Boltzmann weighted sums over configurations with a given cluster size as described in Appendix C.  Lines are drawn between the points as a guide to the eye.
}
\end{figure}

For architectures $\bthree$ and $\bfour$, it is not possible to construct an assembly path for which monomers form multiple bonds at all cluster sizes greater than two (see Fig.~\ref{fig:asspath}).  Therefore, free energy profiles consistent with these architectures, shown in Fig.~\ref{fig:fenergy}, have numerous free energy barriers and local minima.  At parameter values that are optimal for $\bfive$ assembly, progression of $\bfour$ and $\bthree$ intermediates through the assembly cascade is stymied by these free energy barriers. 

All free energy barriers vanish if parameters are increased until formation of single bonds is favorable. Since dimers are stable under these conditions, however, too many capsids initiate and the system becomes mired in the kinetic trap shown in Fig.~\ref{fig:kt}a. At moderate parameter values, $\bfour$ intermediates that correspond to local minima, such as trimers and hexamers, are metastable. Optimal assembly occurs when these intermediates bind to growing capsids as fast as they are formed.  The first local minimum for $\bthree$ capsids does not occur until a cluster size of five; in this case, optimal assembly occurs at parameter values for which dimerization is only slightly unfavorable and binding events involving dimers are common (see Fig.~\ref{fig:nbe}). As parameter values are increased beyond optimal assembly conditions, formation of small intermediates becomes more rapid and multimer binding becomes more important for all capsid designs.

Note that these free energies compare the relative stability of different multimers.  There are also free energy barriers not shown that are associated with subunit binding or unbinding, which is required to transition between these states.  

Given the importance of multimer binding to assembly of some capsid designs, it might seem surprising that a rate equation model \cite{Zlotnick1994,Endres2002} that assumes capsids grow only via addition of monomers predicts significant assembly for some parameter values.  That particular class of models assumes, however, that monomers bind to intermediates with an average binding free energy, independent of the number of intersubunit contacts formed in a particular configuration.  While not explicitly described herein, we have considered a model in which binding free energies were specifically calculated for each capsid intermediate, but the assumption that only monomers could bind to these intermediates was retained. The assembly dynamics predicted by this model for $\bfive$ were consistent with those seen in Brownian dynamics simulations, but concentrations of intermediates at local free energy minima built up for $\bfour$ and $\bthree$ designs, as well as for the design shown in Fig 1B of Ref.~\onlinecite{Endres2002}.  Because these intermediates could not bind with each other, the predicted assembly was much less efficient than that predicted by Brownian dynamics simulations.  These results suggest that it may be important to consider binding of complexes that are larger than the basic assembly unit when hypothesizing assembly mechanisms from experimental data.

\begin{figure}[hbt]
\epsfig{file=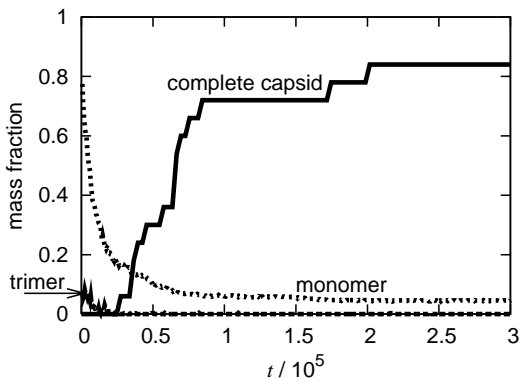,width=.8\linewidth}
\caption{\label{fig:pnnt}
The mass fraction of monomers,  trimers, and complete capsids (60 subunits, with $\nb=4$ bonds each) as a function of time, $t$,  for design $\bfour$, illustrating that intermediate concentrations are always small during successful assembly.  The parameters are those given in Fig~\ref{fig:optass}.  
}
\end{figure}

For many viruses, the basic assembly units are believed to be small intermediates, such as dimers or trimers. Experimentally observed high concentrations of these species during assembly reactions imply that they form rapidly and are extremely metastable. This feature could be included in our model by designing a new ``subunit'' that represents the basic assembly unit, or by choosing higher binding energies for certain bond vectors. In this work, where all maximum binding energies are equal, the importance of multimer binding for $\bfour$ and $\bthree$ capsids is not a result of the interaction between individual subunits. Rather, the collective interactions of many subunits lead to a free energy profile with numerous local minima, which forces assembly to proceed through binding of intermediates. Although binding of trimers and other multimers is essential to $\bfour$ assembly in our simulations, the fraction of subunits that comprise these intermediates is always small compared to the fraction of subunits that are monomers or in completed capsids (see Fig.~\ref{fig:pnnt}).  Hence, the significance of binding of multimers would be difficult to detect by bulk experiments, such as light scattering or size exclusion chromatography, alone.  The combination of these techniques, though, with selective deletion of residues through mutation \cite{Reguera2004} or single molecule experiments may offer insights into the importance of various intermediates in capsid assembly.

\begin{figure}[hbt]
\epsfig{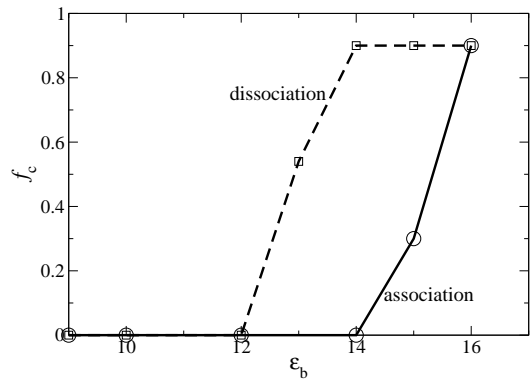}
\caption{\label{fig:dissoc}
Capsid yeilds as a function of $\eb$ (or equivalently inverse temperature or inverse denaturant concentration) illustrating hysteresis between association and dissociation of capsids.  The final capsid yields, $\ct$ at $\tobs=\tf$, are shown for simulations started with subunits in random configurations (association) and simulations started with the final configuration from the $\bthree$ simulation shown in Fig.~\ref{fig:optass}, which had $\ct=0.9$ (dissociation). Parameter values were $\conc=0.11$ and $\tm=0.5$.
}
\end{figure}

{\bf Capsids are metastable in infinite dilution and capsid disassembly shows hysteresis.}
Since even single virions can sometimes infect cells, viral capsids must be metastable in infinite dilution. Model capdsids also displayed this feature; for example, significant assembly with $\conc=0.008$ did not occur for $\eb<20.0$ (see Fig.~\ref{fig:kpd}), but isolated complete capsids simulated without periodic boundary conditions (to model infinite dilution) were metastable through $t=\tf$ for $\eb \ge 14.0$. This result indicates that the final yield of the capsids at some binding energies will be different for a trajectory that started with mostly complete capsids then for a trajectory that started with random subunit configurations, as shown in Fig~\ref{fig:dissoc}. In other words, there would be hysteresis between assembly and disassociation.  

Hysteresis arises because, as discussed above, there can be large free energy barriers separating the early stages of assembly, where there are few bonds per subunit, and complete capsids, for which each subunit has $\nb$ bonds. Therefore, monomers (or complete capsids) can be metastable throughout a finite length trajectory above (or below) $\ccc$. Once the first subunit is removed from a metastable complete capsid, neighboring subunits have fewer bonds and further disassembly is rapid.  Hysteresis in capsid assembly--disassembly has been seen with experiments and theory by Zlotnick and co-workers \cite{Singh2003}. 

\section{Conclusions and outlook}
In this work we examined ensembles of assembly trajectories for models that assemble into capsid-like objects. We demonstrated that computational models can distinguish driving forces and corresponding mechanisms that lead to successful assembly from those that engender dynamic frustration.  Therefore, the approach we have outlined might be used to good effect to analyze the dynamics of other assembly models (e.g., \cite{Schwartz1998, Rapaport2004, Zhang2004, Chen2005}). as well as that for other types of assembly models.  In addition to the calculations presented herein, there are other ways that ensembles of assembly trajectories can be carefully analyzed.  For instance, distributions of capsid formation times could be studied by simulation and potentially estimated with single molecule experiments; comparisons could shed light on assembly mechanisms, such as the mechanisms for $\bfour$ and $\bfive$ assembly illustrated in this work. Trajectories generated by the approach described in this work can be a starting point for performing importance sampling in trajectory space \cite{Dellago1998,Bolhuis2002,E2005}, which would facilitate statistical analysis of ensembles of assembly pathways.

Rechtsman and co-workers describe an ``inverse statistical mechanical-methodology'' that allows importance sampling in model space to design potentials that direct assembly into a particular ground state \cite{Rechtsman2005}.  The ability to generate ensembles of dynamical trajectories invites a related strategy, in which one seeks to optimize a function of entire assembly paths, such as capsid assembly times or identities of key intermediates.  This approach would involve importance sampling steps in trajectory space, such as shooting moves \cite{Dellago2001}, as well as sampling steps in model space.  Understanding model features that lead to specific assembly behavior could guide and interpret experiments that involve mutated capsid proteins.\\

{\bf Acknowledgements} Financial support for this work was provided by NSF, grant CHE0078458, DOE grant CH04CHA01, and MFH was supported by a NIH fellowship, grant F32 GM073424-01.  The authors are indebted to Tracy Hsiao and Matthew Wyndham for assistance in preparing the manuscript and MFH thanks Sander Pronk for helpful discussions and Carlos Bustamante for introduction to this problem.

\appendix
\section{}
\label{sec:appendixa}

The equations of motion given in Eq. \ref{eq:eom} were integrated as follows.  Translational displacements are calculated as described in Eq. 7 of Ref. \onlinecite{Branka1999}.  Bond vector orientations are specified in body fixed coordinates at the beginning of the simulations.  The space fixed coordinates, $\{\mathbf{b}^{(\alpha)}(t)\}$, are determined from a rotation matrix, $\Arot(t)$, which is evolved in time using quaternions, which satisfy the equation of motion given in Eq. 3.37 of Ref. \onlinecite{Allen1987}.  This equation requires angular velocities, $\oomega$, which are determined in an analogous fashion to the translational displacements
\begin{equation}
\oomega_i=\gamr/2\left(\ttau_i^a+\ttau_i^b\right)+\dtau_i
\label{eq:omega}
\end{equation}
where the torques are calculated at two points 
\begin{eqnarray}
\ttau_i^a(t) & \equiv & \ttau_i\left(\{\ri(t),\bbond(t)\}\right) \nonumber \\
\ttau_i^b(t) & \equiv &\ttau_i\left(\{\ri^b(t),\bbond^b(t)\}\right)
\label{eq:taua}
\end{eqnarray}
with the predictor positions determined as in Ref. \onlinecite{Branka1999}
\begin{equation}
\ri^b=\ri+\delta t \left(\gamma\f_i+\df_i\right)
\label{eq:rb}.
\end{equation}
The predictor bond orientations, $\{\bbond^b(t)\}$, are determined from a predictor rotation matrix, which is calculated from Eq. 3.37 of Ref. \onlinecite{Allen1987}, using predictor angular velocities calculated as 
\begin{equation}
\oomega_i^b=\gamr\ttau_i^a+\dtau_i
\label{eq:wb}.
\end{equation}
This formulation assumes that subunits are hydrodynamically isolated and that rotational and translational sources of friction are not coupled; these assumptions can be relaxed as in Ref. \onlinecite{Dickinson1985}.

\section{}
\label{sec:appendixb}

The contribution of multimer-binding to the final assembly product was calculated from simulations as follows. Multimers were designated by clustering subunits connected by one or more bonds. A binding event occurred when the size of a cluster changed, either through a combination of two clusters (positive binding) or division of a cluster (negative binding). Cluster sizes were output every 10 steps; more than one binding event involving the same cluster within 10 steps was found to be exceptionally rare.  The size of a binding event, $\nbe$, was defined by the size of the smallest reactant cluster for positive binding or the smallest product cluster for negative binding.  The net forward binding due to events of size $\nbe$ is given by 
\begin{equation}
\nfb(\nbe)=\nbe \left(\bpls(\nbe)-\bmns(\nbe) \right) 
\label{eq:nbe}
\end{equation}
where $\bpls$ and $\bmns$ are the number of positive and negative binding events of size $\nbe$, respectively. 

\section{}
\label{sec:appendixc}
The free energy, $\Delta g_i$, to build a particular capsid configuration, $i$, from a bath of subunits at concentration $\conc$, can be determined by analogy to Eq.~\ref{eq:sdimer} if the dependence of binding entropy on the number of bonds is neglected

\begin{equation}
\Delta g_i = \sum_{j=1}^{n_i}\left[-b_j\eb/2 \right] -(n_i-1) T\left[\kb \ln(\pi \sigma^3\conc/6) + \sbond(1)\right]
\label{eq:fcapsid},
\end{equation}
where $j$ sums over each subunit in the capsid, $n_i$ is the number of subunits in configuration $i$, and $b_j$ is the number of bonds for subunit $j$. We project the free energy onto the number of particles in a capsid, $n$, by summing over configurations containing $n$ of subunits
\begin{equation}
G_n=- \kt \ln {\displaystyle \sum}_i \mathrm{\delta}_{n_i,n} \exp\left(-\beta \Delta g_i\right) 
\label{eq:Gn}.
\end{equation}
   We performed this sum with Monte Carlo simulations in which trial capsid configurations were generated by adding or deleting subunits from current configurations, and then accepted or rejected according to the Metropolis \cite{Metropolis1953} criterion, with the Boltzmann distribution given by Eq.~\ref{eq:fcapsid}.  Only configurations consistent with minimum energy bonding were considered; thus, this approach was only applicable for parameters with which misformed capsids do not occur.  The average free energy for capsids of size $n$ was efficiently calculated by carrying out umbrella sampling \cite{Chandler1987} in which a harmonic potential as a function of capsid size was used to bias the number of subunits in the capsid.


\begin{thebibliography}{49}
\expandafter\ifx\csname natexlab\endcsname\relax\def\natexlab#1{#1}\fi

\bibitem[{Crick and Watson(1956)}]{Crick1956}
Crick, F. H.~C., and J.~D. Watson. 1956.
\newblock Structure of small viruses.
\newblock \emph{Nature (London)} 177:473--5.

\bibitem[{Caspar and Klug(2004)}]{Caspar1962}
Caspar, D. L.~D., and A.~Klug. 2004.
\newblock Physical principles in the construction of regular viruses.
\newblock \emph{Cold Spring Harbor Symp. Quant. Biol.} 4:1407--1413.

\bibitem[{Bruinsma et~al.(2003)Bruinsma, Gelbart, Reguera, Rudnick, and
  Zandi}]{Bruinsma2003}
Bruinsma, R.~F., W.~M. Gelbart, D.~Reguera, J.~Rudnick, and R.~Zandi. 2003.
\newblock Viral self-assembly as a thermodynamic process.
\newblock \emph{Phys. Rev. Lett.} 90:248101.

\bibitem[{Zandi et~al.(2004)Zandi, Reguera, Bruinsma, Gelbart, and
  Rudnick}]{Zandi2004}
Zandi, R., D.~Reguera, R.~F. Bruinsma, W.~M. Gelbart, and J.~Rudnick. 2004.
\newblock Origin of icosahedral symmetry in viruses.
\newblock \emph{Proc. Natl. Acad. Sci. USA} 101:15556--15560.

\bibitem[{Twarock(2004)}]{Twarock2004}
Twarock, R. 2004.
\newblock A tiling approach to virus capsid assembly explaining a structural
  puzzle in virology.
\newblock \emph{J. Theor. Biol.} 226:477--482.

\bibitem[{Chen et~al.(2005)Chen, Zhang, and Glotzer}]{Chen2005}
Chen, T., Z.~Zhang, and S.~C. Glotzer. 2005.
\newblock A universal precise packing sequence for self-assembled convex
  structures.
\newblock \emph{preprint} .

\bibitem[{Fraenkel-Conrat and Williams(1955)}]{Fraenkel1955}
Fraenkel-Conrat, H., and R.~C. Williams. 1955.
\newblock Reconstitution of active tobacco mosaic virus from its inactive
  protein and nucleic acid components.
\newblock \emph{Proc. Natl. Acad. Sci. USA} 41:690--698.

\bibitem[{Butler and Klug(1978)}]{Butler1978}
Butler, P.~J., and A.~Klug. 1978.
\newblock The assembly of a virus.
\newblock \emph{Sci. Am.} 329:62.

\bibitem[{Klug(1999)}]{Klug1999}
Klug, A. 1999.
\newblock The tobacco mosaic virus particle: structure and assembly.
\newblock \emph{Phil. Trans. R. Soc. Lond. B} 354:531--535.

\bibitem[{Fox et~al.(1994)Fox, Johnson, and Young}]{Fox1994}
Fox, J.~M., J.~E. Johnson, and M.~J. Young. 1994.
\newblock {RNA}/protein interactions in icosahedral virus assembly.
\newblock \emph{Seminars in Virology} 5:51--60.

\bibitem[{Zlotnick et~al.(1996)Zlotnick, Cheng, Conway, Booy, Steven, Stahl,
  and Wingfield}]{Zlotnick1996}
Zlotnick, A., N.~Cheng, J.~F. Conway, F.~P. Booy, A.~C. Steven, S.~J. Stahl,
  and P.~T. Wingfield. 1996.
\newblock Dimorphism of hepatitis {B} virus capsids is strongly influenced by
  the c-terminus of the capsid protein.
\newblock \emph{Biochemistry} 35:7412--7421.

\bibitem[{Zlotnick et~al.(2000)Zlotnick, Aldrich, Johnson, Ceres, and
  Young}]{Zlotnick2000}
Zlotnick, A., R.~Aldrich, J.~M. Johnson, P.~Ceres, and M.~J. Young. 2000.
\newblock Mechanism of capsid assembly for an icosahedral plant virus.
\newblock \emph{Virology} 277:450--456.

\bibitem[{Singh and Zlotnick(2003)}]{Singh2003}
Singh, S., and A.~Zlotnick. 2003.
\newblock Observed hysteresis of virus capsid disassembly is implicit in
  kinetic models of assembly.
\newblock \emph{J. Biol. Chem.} 278:18249--18255.

\bibitem[{Willits et~al.(2003)Willits, Zhao, Olson, Baker, Zlotnick, Johnson,
  Douglas, and Young}]{Willits2003}
Willits, D., X.~Zhao, N.~Olson, T.~S. Baker, A.~Zlotnick, J.~E. Johnson,
  T.~Douglas, and M.~J. Young. 2003.
\newblock Effects of the {C}owpea chlorotic mottle bromovirus $\beta$-hexamer
  structure on virion assembly.
\newblock \emph{Virology} 306:280--288.

\bibitem[{Casini et~al.(2004)Casini, Graham, Heine, Garcea, and Wu}]{Wu2004}
Casini, G.~L., D.~Graham, D.~Heine, R.~L. Garcea, and D.~T. Wu. 2004.
\newblock In vitro papillomavirus capsid assembly analyzed by light scattering.
\newblock \emph{Virology} 325:320--327.

\bibitem[{Valegard et~al.(1997)Valegard, Murray, Stonehouse, van~den Worm,
  Stockey, and Lijlas}]{Valegard1997}
Valegard, K., J.~B. Murray, N.~J. Stonehouse, S.~van~den Worm, P.~G. Stockey,
  and L.~Lijlas. 1997.
\newblock The three-dimensional structures of two complexes betweeen
  recombinant {MS2} capsids and {RNA} operator fragments reveal
  sequence-specific protein-{RNA} interactions.
\newblock \emph{J. Mol. Biol} 270:724--738.

\bibitem[{Zlotnick(1994)}]{Zlotnick1994}
Zlotnick, A. 1994.
\newblock To build a virus capsid. {A}n equilibrium model of the self assembly
  of polyhedral protein complexes.
\newblock \emph{J. Mol. Biol.} 241:59--67.

\bibitem[{Berger et~al.(1994)Berger, Shor, Tucker-{K}ellogg, and
  King}]{Berger1994}
Berger, B., P.~W. Shor, L.~Tucker-{K}ellogg, and J.~King. 1994.
\newblock Local rule-based theory of virus shell assembly.
\newblock \emph{Proc. Natl. Acad. Sci. USA} 91:7732--7736.

\bibitem[{Zlotnick et~al.(1999)Zlotnick, Johnson, Wingfield, Stahl, and
  Endres}]{Zlotnick1999}
Zlotnick, A., J.~M. Johnson, P.~W. Wingfield, S.~J. Stahl, and D.~Endres. 1999.
\newblock A theoretical model successfully identifies features of hepatitis {B}
  virus capsid assembly.
\newblock \emph{Biochemistry} 38:14644--14652.

\bibitem[{Endres and Zlotnick(2002)}]{Endres2002}
Endres, D., and A.~Zlotnick. 2002.
\newblock Model-based analysis of assembly kinetics for virus capsids or other
  spherical polymers.
\newblock \emph{Biophys. J.} 83:1217--1230.

\bibitem[{Reddy et~al.(1998)Reddy, Giesing, Morton, Kumar, Post, Brooks~3rd,
  and Johnson}]{Reddy1998}
Reddy, V., H.~A. Giesing, R.~T. Morton, A.~Kumar, C.~B. Post, C.~L. Brooks~3rd,
  and J.~E. Johnson. 1998.
\newblock Energetics of quasiequivalence: computational analysis of
  protein-protein interactions in icosahedral viruses.
\newblock \emph{Biophys. J.} 1998:546--558.

\bibitem[{Keef et~al.(2005)Keef, Micheletti, and Twarock}]{Keef2005}
Keef, T., C.~Micheletti, and R.~Twarock. 2005.
\newblock Master equation approach to the assembly of viral capsids.
\newblock arXiv:q-bio.BM/0508030 v1.

\bibitem[{Rapaport(2004)}]{Rapaport2004}
Rapaport, D.~C. 2004.
\newblock Self-assembly of polyhedral shells: a molecular dynamics study.
\newblock \emph{Phys. Rev. E.} 70:051905.

\bibitem[{Berger et~al.(2000)Berger, King, Schwartz, and Shor}]{Berger2000}
Berger, B., J.~King, R.~Schwartz, and P.~W. Shor. 2000.
\newblock Local rule mechanism for selecting icosahedral shell geometry.
\newblock \emph{Discrete Applied Mathematics} 104:97--111.

\bibitem[{Schwartz et~al.(2000)Schwartz, Garcea, and Berger}]{Schwartz2000}
Schwartz, R., R.~L. Garcea, and B.~Berger. 2000.
\newblock "{L}ocal rules" theory applied to polyomavirus polymorphic capsid
  assemblies.
\newblock \emph{Virology} 268:461--470.

\bibitem[{Schwartz et~al.(1998)Schwartz, Shor, Prevelige, and
  Berger}]{Schwartz1998}
Schwartz, R., P.~W. Shor, P.~E.~J. Prevelige, and B.~Berger. 1998.
\newblock Local rules simulation of the kinetics of virus capsid self-assembly.
\newblock \emph{Biophys. J.} 75:2626--2636.

\bibitem[{Rapaport et~al.(1999)Rapaport, Johnson, and Skolnick}]{Rapaport1999}
Rapaport, D.~C., J.~E. Johnson, and J.~Skolnick. 1999.
\newblock Supramolecular self-assembly: molecular dynamics modeling of
  polyhedral shell formation.
\newblock \emph{Comput. Phys. Comm.} 121:231--235.

\bibitem[{Zhang and Glotzer(2004)}]{Zhang2004}
Zhang, Z., and S.~C. Glotzer. 2004.
\newblock Self-assembly of patchy particles.
\newblock \emph{Nano Letters} 4:1407--1413.

\bibitem[{Xie and Chapman(1996)}]{Xie1996}
Xie, Q., and M.~S. Chapman. 1996.
\newblock Canine parvovirus capsid structure, analyzed at 2.9 {{\AA}}
  resolution.
\newblock \emph{J. Mol. Biol.} 264:497--520.

\bibitem[{Andersen et~al.(1972)Andersen, Chandler, and Weeks}]{Andersen1972}
Andersen, H.~C., D.~Chandler, and J.~D. Weeks. 1972.
\newblock Roles of repulsive and attractive forces in liquids: the optimized
  random phase approximation.
\newblock \emph{J. Chem. Phys.} 56:3812--3823.

\bibitem[{Allen and Tildesley(1987)}]{Allen1987}
Allen, M.~P., and D.~J. Tildesley. 1987.
\newblock Computer Simulation of Liquids.
\newblock Oxford University Press, New York.

\bibitem[{Branka and Heyes(1999)}]{Branka1999}
Branka, A.~C., and D.~M. Heyes. 1999.
\newblock Algorithms for {B}rownian dynamics computer simulations:
  Multivariable case.
\newblock \emph{Phys. Rev. E.} 60:2381--2387.

\bibitem[{Heyes and Branka(2000)}]{Heyes2000}
Heyes, D.~M., and A.~C. Branka. 2000.
\newblock More efficient {B}rownian dynamics algorithms.
\newblock \emph{Mol. Phys.} 98:1949--1960.

\bibitem[{Chandler(1987)}]{Chandler1987}
Chandler, D. 1987.
\newblock Introduction to Modern Statistical Mechanics.
\newblock Oxford University Press, New York.

\bibitem[{Erickson and Pantaloni(1981)}]{Erickson1981}
Erickson, H.~P., and D.~Pantaloni. 1981.
\newblock The role of subunit entropy in cooperative assembly: Nucleation of
  microtubules and other two-dimensional polymers.
\newblock \emph{Biophys. J.} 34:293--309.

\bibitem[{Ben-tal et~al.(2000)Ben-tal, Honig, Bagdassarian, and
  Ben-Shaul}]{Ben-Tal2000}
Ben-tal, N., B.~Honig, C.~Bagdassarian, and A.~Ben-Shaul. 2000.
\newblock Association entropy in adsorption processes.
\newblock \emph{Biophys. J.} 79:1180--1187.

\bibitem[{A. and Gelbart(1994)}]{Ben-Shaul1994}
A., B.-S., and W.~M. Gelbart. 1994.
\newblock \emph{In} Micelles, Membranes, Microemulsions, and Monolayers, W.~M.
  Gelbart, A.~Ben-Shaul, and D.~Roux, editors, chapter~1. Springer-Verlag, New
  York.

\bibitem[{Maibaum et~al.(2004)Maibaum, Dinner, and Chandler}]{Maibaum2004}
Maibaum, L., A.~R. Dinner, and D.~Chandler. 2004.
\newblock Micelle formation and the hydrophobic effect.
\newblock \emph{J. Phys. Chem. B} 108:6778--6781.

\bibitem[{Kegel and van~der Schoot(2004)}]{Kegel2004}
Kegel, W.~K., and P.~van~der Schoot. 2004.
\newblock Competing hydrophobic and screened-coulomb interactions in hepatitis
  {B} virus capsid assembly.
\newblock \emph{Biophys. J.} 86:3905--3913.

\bibitem[{Ceres and Zlotnick(2002)}]{Ceres2002}
Ceres, P., and A.~Zlotnick. 2002.
\newblock Weak protein-protein interactions are sufficient to drive assembly of
  hepatitis {B} virus capsids.
\newblock \emph{Biochemistry} 41:11525--11531.

\bibitem[{Reguera et~al.(2004)Reguera, Carreira, Riolobos, Almendral, and
  Mateu}]{Reguera2004}
Reguera, J., A.~Carreira, L.~Riolobos, J.~M. Almendral, and M.~G. Mateu. 2004.
\newblock Role of interfacial amino acid residues in assembly, stability, and
  conformation of a spherical virus capsid.
\newblock \emph{Proc. Natl. Acad. Sci. USA} 101:2724--2729.

\bibitem[{Dellago et~al.(1998)Dellago, Bolhuis, Csajka, and
  Chandler}]{Dellago1998}
Dellago, C., P.~G. Bolhuis, F.~S. Csajka, and D.~Chandler. 1998.
\newblock Transition path sampling and the calculation of rate constants.
\newblock \emph{J. Chem. Phys.} 108:1964--1977.

\bibitem[{Bolhuis et~al.(2002)Bolhuis, Chandler, Dellago, and
  Geissler}]{Bolhuis2002}
Bolhuis, P.~G., D.~Chandler, C.~Dellago, and P.~L. Geissler. 2002.
\newblock Transition path sampling: {T}hrowing ropes over rough mountain
  passes, in the dark.
\newblock \emph{Annu. Rev. Phys. Chem.} 53:291--318.

\bibitem[{E et~al.(2005)E, Ren, and Vanden-{E}ijnden}]{E2005}
E, W., W.~Ren, and E.~Vanden-{E}ijnden. 2005.
\newblock Finite temperature string method for the study of rare events.
\newblock \emph{J. Phys. Chem. B} 109:6688--6693.

\bibitem[{Rechtsman et~al.(2005)Rechtsman, Stillinger, and
  Torquato}]{Rechtsman2005}
Rechtsman, M., F.~Stillinger, and S.~Torquato. 2005.
\newblock Optimized interactions for targeted self-assembly: Application to
  honeycomb lattice.
\newblock arXiv:cond-mat/0508030 v1.

\bibitem[{Dellago et~al.(2001)Dellago, Bolhuis, and Geissler}]{Dellago2001}
Dellago, C., P.~G. Bolhuis, and P.~L. Geissler. 2001.
\newblock Transition path sampling.
\newblock \emph{Adv. Chem. Phys.} 123:1--78.

\bibitem[{Dickinson et~al.(1985)Dickinson, Allison, and
  McCammon}]{Dickinson1985}
Dickinson, E., S.~A. Allison, and J.~A. McCammon. 1985.
\newblock {B}rownian dynamics with rotation-translation coupling.
\newblock \emph{J. Chem. Soc., Faraday Trans. 2} 81:59--160.

\bibitem[{Metro\-polis et~al.(1953)Metro\-polis, Rosenbluth, Rosenbluth,
  Teller, and Teller}]{Metropolis1953}
Metro\-polis, N., A.~W. Rosenbluth, M.~N. Rosenbluth, A.~H. Teller, and
  E.~Teller. 1953.
\newblock Equation of state calculations by fast computing machines.
\newblock \emph{J. Chem. Phys.} 21:1087--1092.

\bibitem[{Humphrey et~al.(1996)Humphrey, Dalke, and Schulten}]{Humphrey1996}
Humphrey, W., A.~Dalke, and K.~Schulten. 1996.
\newblock {VMD} -- {V}isual {M}olecular {D}ynamics.
\newblock \emph{Journal of Molecular Graphics} 14:33--38.

\end{thebibliography}

\end{document}